\documentclass[11pt]{article}
\pdfoutput=1
 
\usepackage{amsfonts,amssymb,amsmath,amscd,amsthm,latexsym}
\usepackage{natbib}
\bibpunct{(}{)}{,}{a}{,}{,}
\usepackage[pdftex]{graphicx}
\usepackage{framed}

\newtheorem{alga}{Algorithm}{\scshape}{\sffamily}
\newenvironment{algo}
  {\begin{framed}\begin{alga}\begin{em}\begin{sffamily}}
  {\end{sffamily}\end{em}\end{alga}\end{framed}}

\newtheorem{lemma}{Lemma}
\newtheorem{example}{Example}

\newsavebox{\fmbox}

\newcommand{\btheta}{{\boldsymbol \theta}}
\newcommand{\bTheta}{{\boldsymbol \Theta}}

\newcommand{\by}{\mathbf{y}}

\newcommand{\ds}{\displaystyle}
\newcommand{\vs}{\vspace{0.5cm}}
\renewcommand{\vss}{\vspace{0.25cm}}
\newcommand{\E}{\mathbb{E}}
\newcommand{\V}{\mathbb{V}}

\usepackage{geometry}
\usepackage{setspace} 

\title{Adaptive Multiple Importance Sampling}

\author{
{\sc Jean-Marie Cornuet} \\
Centre de Biologie et Gestion des Populations \\
INRA, Montpellier
\and
{\sc Jean-Michel Marin} \\
Institut de Math\'ematiques et Mod\'elisation de Montpellier, \\
(UMR CNRS 5149), Universit\'e Montpellier 2
\and	
{\sc Antonietta Mira} \\
Department of Economics, University of Lugano, Switzerland
\and
{\sc Christian P.~Robert} \\
Universit\'e Paris Dauphine, CEREMADE, \\ 
IUF, and CREST, Paris \\
}

\date{}

\begin{document}

\maketitle


\begin{abstract}
The Adaptive Multiple Importance Sampling
(AMIS) algorithm is aimed at an optimal recycling of past simulations in an iterated importance sampling scheme. The difference with earlier adaptive
importance sampling implementations like Population Monte Carlo is that the importance weights of all 
simulated values, past as well as present, are 
recomputed at each iteration, following the technique of the deterministic multiple mixture estimator 
of Owen and Zhou (2000). Although the convergence
properties of the algorithm cannot be investigated, we demonstrate through a challenging banana shape target distribution
and a population genetics example that the improvement brought by this technique is substantial.

\end{abstract}

\section{Introduction}

Importance sampling (see for instance \cite{ripley:1987}) is a well-established method used to
overcome the difficulties connected with the complexity of simulating from a target
distribution $\Pi$. Its shortcomings are also well-documented, first and foremost the degradation of its performances
against the dimensionality of the problem. Given an importance distribution $Q$, such that $\Pi$ is 
absolutely continuous with respect to $Q$, importance sampling is based on samples $\by_i\sim Q$.
The corresponding importance weights $\omega_i = \pi(\by_i)/q(\by_i)$ are defined in terms of $\pi(\cdot)$ and $q(\cdot)$, the densities of, respectively,
the target and the importance distributions with respect to the same dominating measure $\nu$.
The distribution of those weights customarily deteriorates as the dimension of $\by_i$ increases ($\by_i$ takes values in $\mathbb{R}^p$).
Since, in practical settings, the fine tuning of the importance distribution against the target is difficult,
alternative Markov chain Monte Carlo approaches have often been advocated as being more 
appropriate for large dimensional problems (see  \cite{robert:casella:2004})
but recent attempts have been made to construct importance functions that automatically 
adapt to the target distribution based on earlier
importance samples \cite[see, e.g.][]{ortiz:kaelbling:2000,liu:liang:wong:2001,pennanen:koivu:2004,
rubinstein:kroese:2004}. Those methods are called adaptive importance sampling 
but they also relate to particle filters \citep{gordon:salmon:smith:1993,doucet:defreitas:gordon:2001}
and sequential Monte Carlo methods \citep{doucet:godsill:andrieu:2001,chopin:2002,delmoral:doucet:jasra:2006}.

There are many different strategies or devising adaptive importance
sampling algorithms.  For instance, the generic Population Monte Carlo 
(PMC) scheme of \cite{cappe:guillin:marin:robert:2004}
can be implemented as the D-kernel \citep{douc:guillin:marin:robert:2007a,douc:guillin:marin:robert:2007b} 
algorithm, whose goal is to fit a mixture of $D$
given kernels to the target in terms of either
minimum variance or minimum Kullback-Leibler divergence. While this
algorithm is shown to converge to the optimal solution (meaning either minimum variance or 
minimum Kullback-Leibler divergence) within the class of D-kernels, it is restrictive to a specific
type of importance distributions that may fail to properly represent the target.

In this paper, we propose a novel perspective to pool together
importance samples from different importance sampling distributions. Those various
importance samples $\by_i^t\sim Q_t$ $(0\le t\le T\,,1\le i\le N_t)$ are associated with
importance weights
\begin{equation}\label{eq:nomix}
\omega_i^t=\pi(\by_i^t)/q_t(\by_i^t)\,,
\end{equation}
where $q_t$ and $\pi$ are proper densities.  While those $T$ samples can be crudely merged by keeping these
original importance weights (Robert and Casella, 2004, Chapter 14), there exists a more refined and 
stabilising alternative called {\em deterministic multiple mixture} due to
\cite{veach:guibas:1995} and popularised by \cite{owen:zhou:2000}.

This alternative solution is similar to the defensive sampling approach of \cite{hesterberg:1995}
in that it modifies the denominator of the  importance weight $\omega_i^t$ from the density value in $\by_i^t$,
$q_t(\by_i^t)$, to a mixture of all the densities that produced the $T$ different samples, namely
\begin{equation}\label{eq:mix}
\frac{1}{\sum_{j=0}^T N_j}\sum_{l=0}^T N_l q_l(\by_i^t)\,,
\end{equation}
resulting in the (so-called deterministic) mixture weight
\begin{equation}\label{eq:amix}
\omega_i^t=\pi(\by_i^t)\bigg/\frac{1}{\sum_{j=0}^T N_j}\,\sum_{l=0}^T N_l q_l(\by_i^t) \,.
\end{equation}
This idea has originally been proposed by \cite{veach:guibas:1995} and is validating by the unbiasedness property
{\footnotesize
\begin{equation}\label{eq:noB}
\E\left[\frac{1}{\sum_{j=0}^T N_j}\sum_{t=0}^T\sum_{i=1}^{N_t} \omega_i^th(\by_i^t)\right] = 
\sum_{t=0}^T N_t \int h(\by) \frac{\pi(\by)}{\sum_{l=0}^T N_l q_l(\by)} q_t(\by) \nu(\text{d}\by) =
\int h(\by) \pi(\by) \nu(\text{d}\by) = \E_\Pi\left[h(\by)\right] \,.
\end{equation}
}

The name {\em deterministic mixture weights} stems from the fact that the weights of the
mixture (\ref{eq:mix}) are neither estimated nor varying over time (which is coherent given that the algorithm
is not sequential).  This is a major difference with the PMC schemes of \cite{douc:guillin:marin:robert:2007a,douc:guillin:marin:robert:2007b}
where the weights of the proposals are optimised against an efficiency criterion like
the Kullback--Leibler divergence. Deterministic mixture is thus misses this adaptive feature and our proposal`
called AMIS (for Adaptive Multiple Importance Sampling) aims at bridging this gap.

When compared with the previous works on multiple mixtures, the novelty in AMIS
is that the family $(Q_t)$ of importance sampling distributions
is constructed sequentially {\em and} adaptively.
This means that the importance sampling distribution used at
each iteration $t$ $(1\le t\le T)$ is derived from the past $t-1$
importance weighted samples. More precisely, at each step $t$, 
\begin{enumerate}
\renewcommand\theenumi{\roman{enumi}}
\item the importance weights of all (present and past) simulated variables 
$\by_i^l$ $(1\le l\le t\,,~1\le i\le N_t)$ are modified, based on the 
current collection of proposals (importance sampling distributions) $(Q_l)_{0\le l\le t}$, and 
\item\label{recy} the entire collection of importance samples partakes to the construction of
the next importance function, $Q_{t+1}$+.
\end{enumerate}

Note that, while \eqref{recy} is a classical feature of Population Monte Carlo algorithms,
most implementations that derive $Q_t$ from past iterations
\citep{douc:guillin:marin:robert:2007b,cappe:douc:guillin:marin:robert:2008} restricted  to use only samples
produced at the previous generation, $t-1$. However, using the entire past of the simulation process provides a natural
stabilisation that speeds up convergence but require 	a much more involved mathematical machinery.
A similar type of methodology has been independently studied by \cite{raftery:bao:2010}.

In most practical settings where importance sampling is implemented, primarily in Bayesian estimation, a self-normalized estimator is used instead,
because the density $\pi$ of the target is known only up to a normalizing constant. In such cases, the importance
weights can be evaluated only up to this normalizing constant and thus need to reweighted by the sum of the weights.
By construction, the self-normalized estimator does not depend on this constant.  In the forecoming examples, we therefore always use the
self-normalized AMIS estimator, even for the benchmark banana shape target considered in \S \ref{sec:banana}. 

The plan of the paper is as follow: we detail the reasons for promoting multiple mixture importance sampling in \S \ref{sec:m&m} and analyse some
associated algorithms in \S \ref{sec:amis}, while discussing their theoretical properties in \S \ref{sec:con}.
The performances of the AMIS algorithm are tested in \S \ref{sec:banana} over a challenging banana shape target
distribution and  in \S \ref{sec:imp} over a realistic population genetic application. We stress that the latter has
motivated the development of the proposed methodology. Indeed, the likelihood of a genetic model most often is
not tractable and regardless of the approximation method used, its derivation involves a non-negligible cost.
We point out that \cite{siren:marttinen:corander:2010} have resorted to our AMIS algorithm to handle complex
population genetics models, avoiding the dramatic consequences of a poor first proposal.

\section{Multiple mixtures}\label{sec:m&m}

The modification in the importance weights from the original ratios \eqref{eq:nomix}
to the mixture ratios \eqref{eq:amix} may sound
surprising or even paradoxical in that the simulated values (and therefore the distributions used to
simulate those) have not changed. We thus detail in this section the motivations for using multiple
mixtures.
There exists a fundamental methodological difficulty in using several importance functions at once. Indeed, if $\Pi$ is the
target density and $Q_0,\ldots,Q_T$ are $T$ different importance functions, samples $\by_1^0,\ldots,\by_{N_0}^0$,
$\ldots$, $\by_1^T,\ldots,\by_{N_T}^T$ that are simulated from these importance functions, with associated standard
importance
weights $\omega_i^t = \pi(\by_i^t)/q_t(\by_i^t)$, can be merged together in that the empirical distribution
function 
$$
\sum_{t,i} \omega_i^t \delta_{\by_i^t}(\by) \bigg/ \sum_{t,i} \omega_i^t
$$
produces in the marginal sense an output approximatively distributed from the target $\pi$. \\
Unfortunately, this property is not sufficient to ensure that the resulting sample
performs satisfactorily. For instance, if one of the importance functions $q_t$ is associated
with an infinite variance in the weights $\omega_i^t$, i.e.~if $\mathbb{E}[(w_i^t)^2]=+\infty$
for one $0\le t\le T$, the potentialy very large weights
resulting from this importance experiment will remain very large in the cumulated sample, no
matter how efficient the other importance functions are. Therefore, the poorly performing sample
will overwhelmingly dominate the other samples in the final approximation and thus ruin the 
overall performances 
of the method. The conclusion of this point is that the raw mixing of importance
samples and of their importance weights, when using different proposals, can be quite harmful, 
when compared with using a single sample, even when most proposals are efficient.

As discussed at large in \cite{owen:zhou:2000}, using a deterministic mixture as a representation
of the production of the simulated sample has the potential to exploit the most efficient proposals
in the sequence $Q_0,\ldots,Q_T$ without rejecting any simulated value nor sample, while reducing the variance
of the corresponding estimators. The poorly performing importance functions are simply eliminated through the erosion of their weights
$$
\pi(\by_i^t) \bigg/ \frac{1}{\sum_{j=0}^T N_j} \sum_{l=0}^T N_l q_l(\by_i^t)
$$
as $T$ increases. Indeed, for all $N_i\geq 1$ not necessarily equals, if $q_0$ is the poorly performing proposal,
while the $q_l$'s $(l>1)$ are good approximations of $\pi$, for a value $\by_i^0$ such that $\pi(\by_i^0)/q_0(\by_i^0)$ is large,
because $q_0(\by_i^0)$ is small,
$\pi(\by_i^0) \big/ \{ N_0 q_0(\by_i^0) + \ldots + N_T q_T(\by_i^0) \}$ 
will behave like $\pi(\by_i^0) \big/ \{ N_1 q_1(\by_i^0) + \ldots + N_T q_T(\by_i^0) \}$ and
decrease to zero as $T$ increases.


\section{The AMIS algorithm}\label{sec:amis}

As explained in the introduction, the idea at the core of the AMIS algorithm is that, for each time-step $t$,
we should update not only the weights $\omega_i^t$ of the $N_t$ current particles, $\by_i^t$,
but also the weights $\omega_i^l$ of all past particles $\by_i^l,\quad 0\leq l\leq t-1$.
Our algorithm can thus be interpreted as a Rao-Blackwell
type of importance sampling where the whole sample of $\sum_{j=0}^T N_j$ 
points can be envisioned of as being homogeneously sampled from 
a deterministic mixture made of the overall sum of proposals. (Once again, the term {\em deterministic mixture}
is a misnomer in that the overall sample is not the outcome of a mixture simulation.)

The major difference with various PMC versions \citep{cappe:guillin:marin:robert:2004,
douc:guillin:marin:robert:2007a,douc:guillin:marin:robert:2007b,
cappe:douc:guillin:marin:robert:2008} is that every single simulated value is recycled and reweighted
at every step of our iterative algorithm by virtue of selecting the appropriate deterministic mixture. 
Indeed, at each iteration $t$ of the algorithm, 
a new adaptive importance sampling distribution is constructed by using, not only the particles
corresponding to the current iteration, but all the weighted particles, based on a well-chosen efficiency criterion
as in earlier PMC versions \citep{cappe:douc:guillin:marin:robert:2008}. In the most standard case when the proposal $Q_t$ is
parameterised, i.e.~when $Q_t$ is of the form $Q(\btheta_t)$ within a parametric family of distributions
$\left\{Q(\btheta),\btheta\in\bTheta\right\}$, the
adaptivity consists in estimating $\btheta_t$ by $\hat\btheta_t$ at each iteration, using all the weighted samples
accumulated so far; this estimation is obtained by using specific criterion like moment matching, variance
minimization or Kullback-Leibler minimization.


A pseudo-code representation of the generic AMIS algorithm is given as follows:

{\footnotesize \singlespacing \begin{algo}\label{algo:genamis}{\bf Generic AMIS}

\vss \noindent At iteration $t=0$, 
\begin{itemize}
\item[\textbf{1)}]  Independently generate 
$N_0$ particles $\by_i^0$ ($1\leq i\leq N_0$) 
from $Q_0$.
\item[\textbf{2)}] For $1\leq i\leq N_0$, compute
$$
\delta_i^0=N_0q_0(\by_i^0)
\quad\text{and}\quad
\omega_i^0=\pi(\by_i^0)\bigg/q_0(\by_i^0)\,.
$$
\item[\textbf{3)}] Compute the importance sampling parameter estimate
$\hat\btheta^0$ of the  parametric family $\left\{Q(\btheta),\btheta\in\bTheta\right\}$ using the weighted particles
$$
\left(\left\{\by_1^0,\omega_1^0\right\},\ldots,\left\{\by_{N_0}^0,\omega_{N_0}^0\right\}\right)
$$
and a well-chosen estimation criterion.
\end{itemize}

\vs \noindent At iteration $t=1,\ldots,T$

\begin{itemize}
\item[\textbf{1)}] Independently generate 
$N_t$ particles $\by_i^t$
($1\leq i\leq N_t$) as 
$
x^t_i \sim Q\left(\hat\btheta^{t-1}\right)\,.
$
\item[\textbf{2)}] For $1\leq i \leq N_t$, compute the multiple mixture at $x^i_t$
$$
\delta_i^t=N_0q_0(\by_i^t)+\sum_{l=1}^{t} N_{l} q\left(\by_i^t;\hat\btheta^{l-1}\right)
$$
and derive the importance weight of particle $\by_i^t$,
$$
\omega_i^t=\pi\left(\by_i^t\right)\bigg/\left[\delta_i^t\bigg/\sum_{j=0}^{t}N_j\right]\,.
$$
\item[\textbf{3)}] For $0\leq l\leq t-1$ and $1\leq i\leq N_l$, update the past importance weights as
$$
\delta_i^l\longleftarrow \delta_i^l+N_lq\left(\by_i^l;\hat\btheta^{t-1}\right)
\qquad\text{and}\qquad
\omega_i^l\longleftarrow {\ds \pi\left(\by_i^l\right)}\bigg/\left[\delta_i^l\bigg/\sum_{j=0}^tN_j\right]\,.
$$
\item[\textbf{4)}] Compute the parameter estimate $\hat\btheta^t$ using all the weighted particles
$$
\left(\left\{\by_1^0,\omega_1^0\right\},\ldots,\left\{\by_{N_0}^0,\omega_{N_0}^0\right\},\ldots,\left\{\by_1^t,\omega_1^t\right\},\ldots,\left\{\by_{N_t}^t,\omega_{N_t}^t\right\}\right)
$$
and the same estimation criterion.
\end{itemize}
\end{algo}
}

After $T$ iterations of the AMIS algorithm, for any $\Pi$-integrable function $h$,
the self-normalized AMIS estimator of $\E_\Pi(h(\by))=\int h(\by)\pi(\by)\nu(\text{d}x)$ is:
\begin{equation}
\label{eq:AMIS}
\widehat{\E_\Pi(h(\by))}=\frac{1}{\sum_{t=0}^T\sum_{i=1}^{N_t} \omega_i^t} \sum_{t=0}^T \sum_{i=1}^{N_t} \omega_i^t h(\by_i^t)\,.
\end{equation}

Since the above algorithm is set in generic terms, we describe a first special case that applies to many
settings and can be seen as a vanilla AMIS algorithm. As in the most recent PMC algorithm of
\cite{cappe:douc:guillin:marin:robert:2008}, the proposal distribution $Q$ is a Student's $t$ proposal,
$\mathcal{T}_3(\mu,\Sigma)$ whose mean $\mu$ and covariance $\Sigma$ parameters are updated by estimating
both first moments of the target distribution $\Pi$ using self-normalized AMIS estimators: 
$\hat\btheta^t=\left(\mu^t,\Sigma^t\right)$ and
\begin{equation}
\hat\mu^t=\dfrac{\sum_{l=0}^t\sum_{i=1}^{N_l}\omega_i^l \by_i^l }{\sum_{l=0}^t\sum_{i=1}^{N_l}\omega_i^l}
\quad\text{and}\quad
\hat\Sigma^t=\dfrac{\sum_{l=0}^t\sum_{i=1}^{N_l}\omega_i^l(\by_i^l-\hat\mu^t)(\by_i^l-\hat\mu^t)^\text{T} }{
\sum_{l=0}^t\sum_{i=1}^{N_l}\omega_i^l}\,.
\label{eq:studentAMIS}
\end{equation}
Note that the degrees of freedom of the $t$ distribution are always set to $3$ as the lowest value 
allowing for finite first moments but they could also be estimated at each iteration.
Moreover, instead of using the previous ``moments matching'' criterion, we can also used the Kullback-Leibler divergence
between $\Pi$ and $Q$ in order to choose the parameter $\btheta=(\mu,\Sigma)$,
$$
\text{div}(\Pi,Q(\btheta)) = \int \log\dfrac{\pi(\by)}{q(x;\btheta)}\,\pi(\by)\,\nu(\text{d}\by)\,.
$$
Here, the best choice for the parameter $\btheta$ is the maximum likelihood estimate of
$\btheta$ where the observations are weighted by their corresponding importance weight.
These two different strategies give essentially the same results.

Quite obviously and as illustrated by the next section, more elaborate proposals
are possible, depending on the information available on $\Pi$. 
For instance, if the potential for multimodality of the target $\Pi$ is high enough, a mixture 
of Student's $t$ distributions as in \cite{cappe:douc:guillin:marin:robert:2008} would be more appropriate. 
When dealing with a Bayesian hierarchical model, creating classes (or blocks) of components of the parameter
in agreement
with the hierarchical levels (as in Gibbs sampling) and designing the Student's $t$ proposals
block by block should also be more efficient. 


Similarly, matching the expectation and the covariance structure of
the Student's proposal distribution with both first moments of the target distribution is only one among
many efficiency criteria that can be used to calibrate the parameters of the proposal distribution at 
each step of the algorithm. For instance, as done in the next section, we can alternatively minimise 
the Kullback-Leibler divergence between the target and the proposal distribution 
following the approach of \cite{cappe:douc:guillin:marin:robert:2008}.

Although we do not elaborate on this possible improvement, note also that, once the weighted sample based on
$\sum_{t=0}^T N_t$ simulations is obtained, it is possible to apply a final clustering 
(standard) algorithm on this sample, based on a Gaussian mixture representation.
Those clusters can be used to estimate local covariance and mean
structures and then  simulate a final and global sample based on the
cluster representation but using Student's $t$ distributions.
Because all weights are controlled, we can then merge this final sample with the sequence of
earlier samples without losing the deterministic representation.

A special version of interest of the AMIS algorithm is based on the used of mixtures of 
multivariate Gaussian densities. That is
$$
q(\by;\btheta) = \sum_{i=1}^k \rho_i \varphi(\by;\mu_i,\Sigma_i)\,,
\qquad
\sum_{i=1}^k \rho_i =1\,,
$$
where $\varphi(\cdot;\mu,\Sigma)$ denotes a multivariate Gaussian density with mean $\mu$ and
covariance matrix $\Sigma$, as in the $D$-kernel approach to PMC algorithms of \cite{cappe:douc:guillin:marin:robert:2008}.
We also use the Kullback-Leibler divergence between $\Pi$ and $Q$ in order to choose the parameter
$\btheta=(\rho_1,\ldots,\rho_k,\mu_1,\ldots,\mu_k,\Sigma_1,\ldots,\Sigma_k)$,
$$
\text{div}(\Pi,Q(\btheta)) = \int \log\dfrac{\pi(\by)}{q(\by;\btheta)}\,\pi(\by)\,\nu(\text{d}\by)\,.
$$
As already mentioned, the best choice for the parameter $\btheta$ is then the maximum likelihood estimate of
$\btheta$. In the AMIS setting, the observations are weighted by their corresponding importance weight:
at iteration $t$ the whole sequence of samples $\by_i^l$ $(0\le l\le t)$ with their updated weights $\omega_i^l$ is used
inside a weighted EM algorithm, which is solved using the {\sf mixmod} software \citep{biernacki:celeux:govaert:langrognet:2006}. 
The number $k$ of components used for the mixture can be either set in advance or, more realistically, 
estimated at iteration $t=0$ by the ICL criterion of \cite{biernacki:celeux:govaert:2000}
and a substantial number $N_0$ of iterations. We do not reproduce the earlier pseudo-codes
for this special case since the differences are minimal. Note that the extension to a mixture of
$t$ densities is equally feasible since there exists a corresponding EM algorithm \citep{Peel:McLachlan:2000}.

\section{Initialization}

A primary difficulty with adaptive importance algorithms is that the starting distribution has a major impact
on the resulting performances of those algorithms.  Due to the ``what-you-get-is-what-you-see'' nature of such
algorithms, it is quite difficult to recover from a poor starting sample, the adaptivity focussing only on the
visited parts of the simulation space. Therefore, we strongly require that a significant part of the computing
effort be spent on the initialization stage.

In order to calibrate this computing effort, we use the effective sample size (ESS).  For a sample of size $N_0$
based on the importance distribution $Q_0$, the ESS is defined by $\ds
\frac{N_0}{1+\V_{Q_0}[\pi(\by)/q_0(\by)]}$ \citep{hesterberg:1995,liu:2001} and it corresponds to the size of
an equivalent iid sample simulated from $\Pi$.  This measure of efficiency does not depend on $h$ and, in
practice,
$$
\V_{Q_0}[\pi(\by)/q_0(\by)]=\int \left\{\pi(\by)/q_0(\by)-\E_{Q_0}[\pi(\by)/q_0(\by)]\right\}^2q_0(\by)\nu(\text{d}\by)
$$ 
can be estimated using the coefficient of variation of the importance weights. 

The initialization solution we propose proceeds through two steps:
\begin{itemize}
\item[1.] Independently generate $N_0$ uniform samples on the $p$-dimensional hyper-cube, \\ $U_1,\cdots,U_{N_0}$ ;
\item[2.] Use an inverse logistic transformation, with scale parameter $s$, to map these points on $\mathbb{R}^p$, independently for each coordinate ;
\item[3.] Maximize the ESS of the points with respect to the scale parameter of the logistic distribution and obtain $s^*$ ;
\item[4.] Use as starting cohort of particles $F^{-1}(U_i;s^*)$, $i=1,\cdots,N_0$ where \\
$F^{-1}(u;s) = s \log(u/(1-u))$ (vectorized expression $u\in\mathbb{R}^p$).
\end{itemize}

Note that to maximize the ESS is equivalent to minimize the variance of the importance weights. Moreover, in
order to maximize the ESS, we only need simulate a single logistic sample since we can adapt the scale of this
sample by mere multiplication. Obviously, this solution is far from fool-proof and we favour an informed
alternative implementation provided items of information on the target distribution are available. Those items
may obviously be provided by multiple pilot runs.

Nelder and Mead's (\citeyear{Nelder:Mead:1965}) algorithm is used to maximize the ESS.  This simplex method
depends on the comparison of the ESS values at the $p+1$ vertices of a general simplex, followed by the
replacement of the vertex with the smallest value by another point.  The simplex keeps adapting to the local
landscape and converges to the global maximum.

\section{Convergence issues and tuning}\label{sec:con}

While establishing unbiasedness and convergence of the deterministic mixture estimator of \cite{owen:zhou:2000}
is relatively straightforward, the introduction of an adaptive mechanism in the construction of the sequence of
proposals highly complicates handling both properties. First, the estimator is no longer unbiased and its
convergence (in $T$ for a fixed values of $N_t$) cannot be established without imposing compactness
restrictions on the simulation space or upper bounds on the target density. 

In order to detail convergence difficulties, we concentrate on the Student's $t$ version of the AMIS algorithm.
Furthermore, we only consider the extreme case $N_t=1$, meaning that each iteration of the algorithm only
processes a single new simulated value: the proposal is then updated after each new iteration. We also simplify
the update of the parameters of the Student's $t$ proposal by restricting learning to the mean $\hat\mu^t$, the
covariance matrix being set to an arbitrary value. This clearly is a formalised setting that we do not advocate in
practice.

The density of the Student's $t$ distribution with 3 degrees of freedom and mean $mu$ is
denoted $t_3(\by;\mu)$. The update of $\mu$ after iteration $t$ is then
$$
\hat\mu^t=u_{t+1}(\by_{0:t})=
\sum_{k=0}^t\frac{\pi(\by_k)\by_k}{q_0(\by_k)+\sum_{i=1}^{t}t_3(\by_k;u_i(\by_{0:i-1}))}\,,
$$
where $u_1(\by_0)={\pi(\by_0)\by_0}\big/{q_0(\by_0)} = \hat{\mu}^0$.

First, the unbiasedness of the estimator $\hat\mu^t$ for every $t>1$ does not
follow from the arguments found in the original version of \cite{owen:zhou:2000} because of
the dependence of the importance weight of $\by_t$ on subsequent $\by_j$'s $(j>t)$.
Indeed, for $t\ge 1$, we have
\begin{align*}
\mathbb{E}[\hat\mu^t] &= \sum_{k=0}^t\mathbb{E}\left[
\frac{\pi(\by_k)\by_k}{q_0(\by_k)+\sum_{i=1}^{t}t_3(\by_k;u_i(\by_{0:i-1}))}\right] \\
&= \sum_{k=0}^t \int \frac{\pi(\by_k)\by_k}{q_0(\by_k)+\sum_{i=1}^{t}t_3(\by_k;u_i(\by_{0:i-1}))}
	t_3(\by_k;u_k(\by_{0:k-1})) \,\text{d}\by_k\\
&\qquad\qquad \times\,g_k(\by_{0:k-1}) \,\text{d}\by_{0:k-1} h_k(\by_{k+1:t}|\by_{0:k})\,\text{d}\by_{k+1:t}
\end{align*}
where $g_k(\by_{0:k-1})$ is the joint distribution of the past simulations and $h_k(\by_{k+1:t}|\by_{0:k})$ is
the conditional distribution of the future simulations given the current and past ones. Due to this latter
term, the full conditional distribution of $\by_k$ given the past and future simulations $\by_{0:k-1}$ and
$\by_{k+1:t}$ is no longer $ t_3(\by_k;u_k(\by_{0:k-1}))$ and this modification implies that $\hat\mu^t$ is
biased. Furthermore, the dependence of this bias on $t$ is so intricate that we cannot manage the asymptotic
bias. A similar impossibility occurs when studying the variance, hence preventing a theoretical conclusion
about the convergence properties of the AMIS algorithm.  Moreover, the standard convergence results on
triangular arrays \citep{douc:moulines:2008} do not apply here, contrariwise to the PMC algorithm
\citep{douc:guillin:marin:robert:2007a}.  Note that a simple if artificial modification of the AMIS
algorithm brings a straightwforward solution to the bias difficulty: when using an additional simulation 
thread for the calibration of the proposal distributions, the arguments of \cite{owen:zhou:2000} apply by
first conditioning upon this second series.

Using exactly the same results as in \cite{douc:guillin:marin:robert:2007a}, notably Theorem A.1 on the convergence
of triangular arrays, under very weak conditions, we obtain the following lemma for the AMIS algorithm:

\begin{lemma}
When $T$ and $N_0,\ldots,N_{T-1}$ are fixed, the estimator $\widehat{\E_\Pi(h(\by))}$ is converging in probability
to ${\E_\Pi(h(\by))}$ when $N_T$ goes to infinity.
\end{lemma}

Indeed, the fact that all sample sizes $N_t$ but the last one $N_T$ are set to a given value means that the
weights of the terms $\by_i^t$ $(0\le t\le T-1)$ converge to $0$, while the bias in the weights of the
$\by_i^T$ asymptotically vanishes, conditionally on the past samples.
The weak conditions are related to the tail behaviour of the importance densities with respect to the target.

However, this setting is not the one in which AMIS should be used. The number of iterations $T$ and the numbers
of simulations $N_t$ ($t=1,\ldots,T$) should be related to the dimension $p$ of the target distribution.  We
recommend to use $N_t$ in the range $25-500$: from $25$ when $p$ is small (typically $d=1$ or $d=2$) to $500$
when $p$ is large (typically $p=20$). We tested this strategy for the AMIS algorithm on various target
distributions. Note we used the default calibration $N_1=N_2=\ldots=N_T$. However, it should be more
efficient to increase the numbers of simulations with the accuracy of the proposal distributions,
$N_1<N_2<\ldots<N_T$, provided an automated scheme on the choice of the $N_t$ can be found.

For instance, in the area of population genetics, \cite{siren:marttinen:corander:2010} proposed an original
Bayesian method for inferring population histories from unlinked single-nucleotide polymorphism. It is used on
an approximation to the neutral Wright-Fisher diffusion that models random fluctuations in allele frequencies.
Inference about the tree topology imply that the posterior distribution be marginalized over a drift parameter,
which, for $K$ populations, is a positive vector of dimension $2K-2$.  \cite{siren:marttinen:corander:2010}
circumvented this difficulty by resorting to the AMIS algorithm.  They used a product of independent Beta
distributions as the initial importance distribution ($Q_0$), then the following importance distributions
($q_t,\quad t=1,\ldots,T$) were defined as multivariate Student's distributions whose parameters are adapted at
fixed interval. In their tests, they chose $N_0=\sum_{t=0}^T N_t/2$ and $T=10-200$ depending on $\sum_{t=0}^T
N_t$. Typically, for simulated datasets where $K=5$, they used $T=200$ and $N_t=50$ and, for an analysis of
human data where $K=7$, they computed the posterior probability of two topologies with $\sum_{t=0}^T
N_t=30,000$.

In connection with the dependence of the simulation numbers $N_t$ on the dimension $p$ of the target
distribution, we note that the AMIS algorithm caters to highly different goals:
\begin{itemize}
\item To compute a numerical approximation of the expectation of a fixed function $h$, 
$I=\int h(x)\Pi(\text{d}x)$ ; 
\item To obtain an approximation of the marginal of a joint distribution;
\item To provide a global approximation of a sample from the target distribution $\Pi$.
\end{itemize}
Depending on the purpose for which the AMIS algorithm is used, the requested minimal value for the ESS will
vary. For instance, if we want to approximate $I$ for a specific function $h$,
then the minimal value for the ESS depends on  $\int (h(x)-\pi(h))^2\Pi(\text{d}x)$.
If, instead, the goal is to approximate the target distribution $\Pi$, the minimal value
for the ESS could be derived from the $L_2$ non-parametric estimation error, that is,
$\int(\hat \pi-\pi)\mu(\text{d}x)$ where $\hat\pi$ is a kernel density approximation of $\pi$. 
As a stopping rule, we propose to iterate the AMIS algorithm, i.e.~to increase $T$, until the desired ESS is
achieved.

The goal of the next Section is to illustrate the fact that the AMIS algorithm can outperform a standard
adaptive importance sampling solution on a benchmark target distributions. That is, with the same adaptive
scheme, this algorith manages to get a significant improvement by pooling together all the simulated points in
the sequential multiple mixture. Given that standard importance sampling algorithms perform well only when $T$
is small and when $N_t$ is large, we also implemented the AMIS algorithm in this setup and chose
$N_0=100,000$, $T=10$ and $N_1=\ldots=N_{10}=10,000$.

In Section \ref{sec:imp}, we instead consider a realistic population genetics example where the AMIS algorithm
is implemented with $N_0=200,000$, $T=2$ and $N_1=N_2=200,000$. The initial importance sampling distribution
$Q_0$ is then the prior distribution.  No optimization procedure related with the ESS is required in this case.
Indeed, for such a target distribution, the region of relevance within the parameter space is easily reached
and we do not need many adaptation steps. However, the calculation of the target density is quite expensive and 
an involved recycling of the whole set of simulations is then relevant.

\section{A banana shape target example}\label{sec:banana}

This evaluation of the performances of AMIS resorts to the benchmark target density of
\cite{haario:saksman:tamminen:1999,haario:saksman:tamminen:2001},  which can be calibrated as to become
extremely challenging. The target density is based on a centered $p$-multivariate Gaussian,
$\by\sim\mathcal{N}_p(0_p,\Sigma)$ with covariance matrix $\Sigma=\text{diag} (\sigma^2,1,\ldots,1)$ which is
twisted by a change of variable in the second coordinate from $y_2$ to  $y_2-b(y_1^2-\sigma^2)$.  Other
coordinates remain unchanged. This change of variable leads to a twisted (or banana shaped) distribution that
has expectation equal to 0 and uncorrelated components. Since the Jacobian of the twisting  transformation is
equal to $1$1, the target density is
$$
\pi(\by)=f_{\mathcal{N}(0_p,\Sigma)}\left(y_1,y_2+b(y_1^2-\sigma^2),y_3,\ldots,y_p\right)\,,
$$
where $f_{\mathcal{N}(0_p,\Sigma)}(\cdot)$ denotes the density of the  centered $p$-multivariate
Gaussian distribution with covariance $\Sigma$. One of the appeals of this benchmark is to allow for various
degrees of heavy tails through the choice of the parameter $b$. 

In this example, we only consider a mild banana shape density, with $\sigma^2=100$ and $b=0.03$. 
More twisted distributions, i.e.~ones with fatter tails, can be obtained by calling for higher values of $b$ and/or $\sigma^2$.
In this case, the target distribution satisfies $\E(y_i)=0$ for all $i=1,\ldots,p $, $\V(y_1)=100$,  $\V(y_2)=19$, and 
$\V(y_i)=1$ for all $i=3,\ldots,p$.


For this target, we compare an iterative importance sampling algorithm that uses the classical mixture version
(as opposed to the deterministic mixture version) with the Gaussian mixture version of the AMIS algorithm.
This reference algorithm, called AIS (for Adaptive Importance Sampling), thus also relies on past simulations
for creating a new Gaussian mixture proposal, but it relies on usual importance weights.  Given the recent work
on PMC algorithms \citep{cappe:douc:guillin:marin:robert:2008}, this can be considered as a state-of-the-art
methodology for the comparison.

For both schemes, an initial sample of $N_0=10^5$ particles is simulated from a standard logistic distribution
and rescaled component-wise to ensure a maximal ESS. In the following, $T=10$ iterations and $N_t=10,000$
particles $(1\le t\le T)$ are used.  

The clustering step fitting a mixture to the weighted samples is solved
via the {\sf mixmod} software \citep{biernacki:celeux:govaert:langrognet:2006}, with the number of components
in the mixture being calibrated via the ICL criterion \citep{biernacki:celeux:govaert:2000} during the first
iteration. It suggested resorting to a mixture of 4 components to correctly fit the banana shape target in two
dimensions. Both schemes under comparison take approximatively the same computing time (depending of course on
the dimension $p$ of the problem) and produce $2\times 10^5$ weighted particles.  Note that, for $p=20$, to
maximize the ESS using the Nelder and Mead algorithm in the initialization step takes almost the same amount of
time than $T=10$ iterations of the AMIS algorithm with $10,000$ particles per iteration.

\begin{table}
\begin{center}
\begin{tabular}{|c||c||c||c|}
\hline
Target   function    & $p$ & AMIS & AIS \\
\hline
\hline
             &  5 & 0.00430 (0.00319) & 0.00473 (0.00664) \\
$\E(y_1)=0$  & 10 & 0.00408 (0.00469) & 0.01221 (0.01224) \\
             & 20 & 0.00840 (0.00875) & 0.03208 (0.03208) \\
\hline
\hline
             &  5 & 0.01044 (0.01486) & 0.01342 (0.01275) \\
$\E(y_2)=0$  & 10 & 0.04589 (0.04419) & 0.05088 (0.03632) \\
             & 20 & 0.06409 (0.02552) & 0.08461 (0.05381) \\
\hline
\hline
$\sum_{l=3}^5\E(y_i)=0$    & 5  & 0.00002 (0.00003) & 0.00009 (0.00008) \\
$\sum_{l=3}^{10}\E(y_i)=0$ & 10 & 0.00009 (0.00014) & 0.00044 (0.00074) \\
$\sum_{l=3}^{20}\E(y_i)=0$ & 20 & 0.00028 (0.00053) & 0.00177 (0.00343) \\
\hline
\hline
               &  5 & 6.795002  (6.72701) & 15.41744 (14.34075) \\
$\V(y_1)=100$  & 10 & 49.94052 (34.21143) & 56.08176 (38.46109) \\
               & 20 & 67.24332 (47.74095) & 94.42488 (58.44744) \\
\hline
\hline
              &  5 &  4.43871  (4.11778) &  8.76941  (6.90886) \\
$\V(y_2)=19$  & 10 & 14.18724  (6.54468) & 25.85457 (12.67837) \\
              & 20 & 23.56200 (12.61588) & 35.76413 (15.90980) \\
\hline
\hline
$\sum_{l=3}^5\V(y_i)=3$     &  5 & 0.00004 (0.00003) & 0.00014 (0.00019) \\
$\sum_{l=3}^{10}\V(y_i)=8$  & 10 & 0.00019 (0.00034) & 0.00069 (0.00104) \\
$\sum_{l=3}^{20}\V(y_i)=18$ & 20 & 0.00212 (0.00245) & 0.00413 (0.00613) \\
\hline
\end{tabular}
\end{center}
\caption{Mean square errors calculated over 10 replications of the AMIS
and AIS schemes for different target functions for different values of $p$ and in
parenthesis the corresponding standard errors.}
\label{tab:rmse}
\end{table}

\begin{figure}[htb]
\centerline{\includegraphics[scale=0.8]{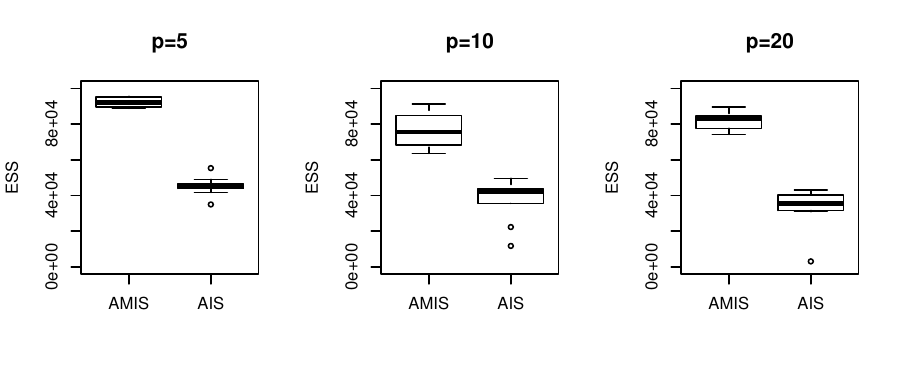}}
\caption{Banana shape example: boxplots of the 10 replicate ESS's for  
the AMIS scheme (left) and
the AIS scheme (right) for $p=5,10,20$. The total number of particles is equal to $200,000$.}
\label{fig:banana-ess}
\end{figure}

\begin{figure}[htb]
\centerline{\includegraphics[scale=0.8]{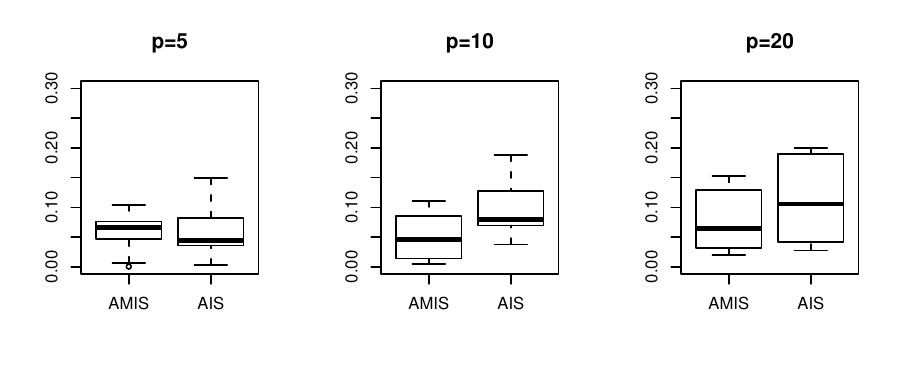}} \vglue -0.5cm
\centerline{\includegraphics[scale=0.8]{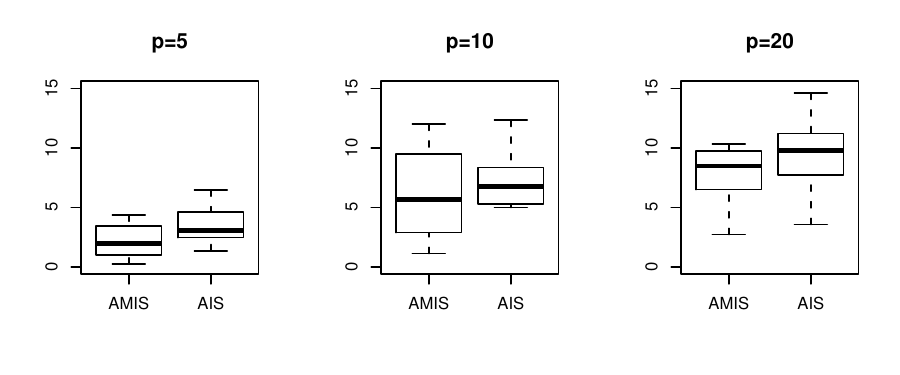}} \vglue -0.5cm
\caption{Banana shape example: boxplots of the 10 replicate absolute  
errors associated to the estimations of $\E(y_1)$ (first line) and $\V(y_1)$ (second line)  
obtained by the AMIS and AIS schemes for $p=5,10,20$.}
\label{fig:banana-mean}
\end{figure}


The results of this experiment are reported in Table \ref{tab:rmse} and on Figures
\ref{fig:banana-ess} and \ref{fig:banana-mean}. These results all are consistent with a domination of the AMIS
scheme. The gain in ESS is quite spectacular, but resulting from the strong stabilisation brought by the AMIS
averaging.  The improvement in root mean square error shown in Table \ref{tab:rmse} typically varies with the
target function as well as with the overall dimension $p$, but may go as far as a threefold reduction. The
boxplots of the absolute errors convey the same message of a uniform domination by AMIS in this setting.

\section{An example from population genetics}\label{sec:imp}

Another illustration of the potential advantage in using the AMIS algorithm is now discussed. It addresses a
realistic population genetics problem that essentially amounts to estimate parameters of an evolutionary
scenario in which two populations have diverged from a common and unknown ancestral population. Data consists
in the genotypes at a single microsatellite locus of 50 diploid individuals sampled from each population. This
locus is assumed to evolve according to the strict Stepwise Mutation model (SMM), i.e., when a mutation occurs,
the number of repeats of the mutated gene increases or decreases by one unit with equal probability. After
divergence, we also assume that populations do not exchange genes (no migration). The four parameters to
estimate are the three effective population sizes ($n_1, n_2$, $n_{Anc}$) and the time of divergence
($t_{div}$), all scaled by the mutation rate ($\mu$) of the locus : $\theta_1$ (=4$n_1\mu$), $\theta_2$
(=4$n_2\mu$), $ \theta_A$ (=4$n_{Anc}\mu$) and $\tau$ (=$t_{div}\mu$). The likelihood of this model is costly
to obtain, which is why we selected this benchmark example. In a Bayesian framework, uniform priors
$\mathcal{U}[0.1,100]$ and $\mathcal{U}[0.005,5]$ were chosen for the parameters $\theta$ and $\tau$,
respectively. Our target is the posterior distribution of $(\theta_1,\theta_2,\theta_A,\tau)$.

Five data files have been simulated with the software \emph{DIYABC}  \citep{cornuetetal:2008}, with the
following parameter values:  $n_1=n_{Anc}=10,000$, $n_2=2,000$, $t_{div}=1,000$, and $\mu=0.0005$, leading
to $\theta_1=\theta_A=20$, $\theta_2=4$, and $\tau=0.5$.  Each dataset has been processed twice.

The first analysis, used as a control, is based  on an MCMC run in which the gene tree of the sampled genes is
updated  together with the four demographic parameters. This has been performed with the  software \emph{IM}
\citep{hey:nielsen:2004}. 

The second analysis combines the \emph{AMIS} algorithm and an estimation of the likelihood based on importance
sampling (IS)  for gene genealogies in the same way as \cite{beaumont:2003} embedded  an IS computation of the
likelihood in a MCMC exploration of the parameter space. We  note that the likelihood of a set of demographic
parameters is  computed by averaging importance weights of gene trees simulated event by event according to
proposal  distributions and parameter values. Each gene tree is built in three  steps looking backward in time:
i) between present time and time of divergence, lineages are coalesced  or mutated following Stephens and
Donnelly's algorithm (\citeyear{stephens:donnelly:2000}), monitoring times of events as in
\cite{beaumont:2003}, ii) at time of  divergence, remaining lineages of both populations are merged and iii)
after divergence, the gene tree is completed according to the \emph{SDPAC} algorithm of
\cite{cornuet:beaumont:2007}. 

To assess the stability of the approximations provided by both  methods, each analysis was repeated four times
(i.e., with four different groups of random seeds for each dataset). Each MCMC (\emph{IM}) was run as a single
chain of $10^7$ updates after a burn-in period of $10^6$ updates.  The IS-AMIS algorithm was run with
$N_0=200,000$, $T=2$, and $N_0=N_1=200,000$. No optimisation based on the ESS was required towards the
calibration of the initial importance function: the prior distribution was deemed satisfactory. Indeed, the
prior distribution is then sufficiently concentrated that there is no difficulty in finding the relevant region
in the parameter space.

\begin{figure}[htbp]
\centerline{\includegraphics[scale=0.4]{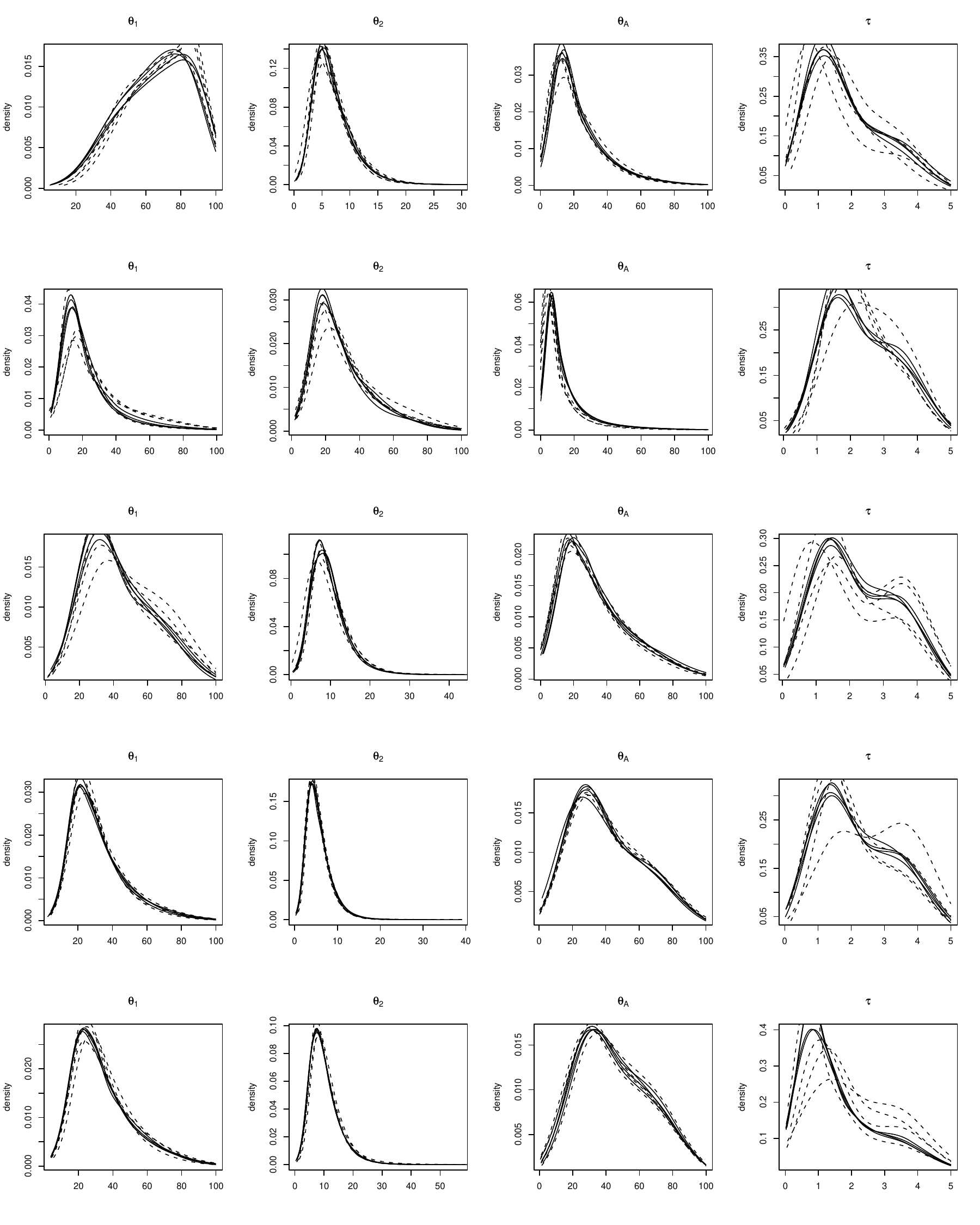}}
\caption{Population genetics example: posterior distributions of the  
four parameters $(\theta_1,\theta_2,\theta_A,\tau)$ for 5 simulated  
datasets obtained
through IS-AMIS (continuous line) and MCMC (dashed line). Each  
analysis has been repeated four times to evaluate the impact of  
repeatability.}
\label{fig:popgen}
\end{figure}

Both methods provided similar outputs as shown on Figure  \ref{fig:popgen}, thus validating the IS-AMIS
approach. However the  major conclusion of this study is that, whereas each MCMC run lasted about 2 hours, the
IS-AMIS executation lasted only approximately 20 min with a slightly better repeatability in that MCMC outputs
were often more  variable. We stress that the calculation of the likelihood function of those models has a
non-negligible cost. We used here an importance sampling  approximation as in \cite{stephens:donnelly:2000} and
the cost of this approximation increases considerably with the number of simulated gene trees. This type of
models is then adequate for the adoption and the development of the AMIS algorithm: all particles simulated
during the process are recycled, which minimizes the number of calls to the likelihood function.  Due to this
recycling process, the AMIS algorithm cannot be easily compared with other adaptive importance sampling schemes
since those do not naturally involve any recycling step and since the natural mixture of importance samples is
fraught with dangers, as explained at the beginning of this paper.

\section{Conclusion}

We have investigated in this paper an adaptive importance sampling method that extends the scope of the
original deterministic multiple mixture approach of \cite{owen:zhou:2000} in that the sequence of importance
proposals sequentially builds on the samples produced so far. The generality of the AMIS algorithm is that it offers a
super-efficiency compared with other adaptive importance sampling techniques by allowing for an integral
recycling of the past simulations. It thus provides a scope for processing those heterogeneous simulations as a
whole and for treating the computing cost $\sum_{t=0}^T N_t$ as a single entity. The challenging issue of the theoretical
convergence of the AMIS algorithm has not been solved in this paper and the most promising direction in
this respect is to derive acceptable growth rates for the sizes $N_t$ when $t$ goes to infinity.

\section*{Acknowledgements}

The authors are grateful to many colleagues for helpful discussions on the convergence properties of the AMIS
algorithm, in particular to  Olivier Capp\'e, Pierre Del Moral, Randal Douc, Nadia Oujdane, Judith Rousseau,
and Vivek Roy.  The authors wish to thank the Associate Editor for its encouraging comments and three reviewers
whose suggestions were very helpful in improving the presentation of this work.  This work has been partly
supported by the Agence Nationale de la Recherche (ANR, 212, rue de Bercy 75012 Paris) through the 2005-2008
projects {\sf Adap'MC} and {\sf Misgepop}, and the 2009-2012 projects {\sf Big'MC} and {\sf Emile}..

\bibliography{CMMR11}

\begin{thebibliography}{}

\bibitem[Beaumont, 2003]{beaumont:2003}
Beaumont, M. (2003).
\newblock Estimation of population growth or decline in genetically monitored
  populations.
\newblock {\em Genetics}, 164:1139--1160.

\bibitem[Biernacki et~al., 2000]{biernacki:celeux:govaert:2000}
Biernacki, C., Celeux, G., and Govaert, G. (2000).
\newblock Assessing a mixture model for clustering with the integrated
  completed likelihood.
\newblock {\em IEEE Transactions on Pattern Analysis and Machine Intelligence},
  22:719--725.

\bibitem[Biernacki et~al., 2006]{biernacki:celeux:govaert:langrognet:2006}
Biernacki, C., Celeux, G., Govaert, G., and Langrognet, F. (2006).
\newblock Model-based cluster and discriminant analysis with the mixmod
  software.
\newblock {\em Computational Statistics and Data Analysis}, 51:587--600.

\bibitem[Capp\'e et~al., 2008]{cappe:douc:guillin:marin:robert:2008}
Capp\'e, O., Douc, R., Guillin, A., Marin, J.-M., and Robert, C. (2008).
\newblock Adaptive importance sampling in general mixture classes.
\newblock {\em Statistics and Computing}, 18:587--600.

\bibitem[Capp\'e et~al., 2004]{cappe:guillin:marin:robert:2004}
Capp\'e, O., Guillin, A., Marin, J.-M., and Robert, C. (2004).
\newblock Population {M}onte {C}arlo.
\newblock {\em Journal of Computational and Graphical Statistics}, 13:907--929.

\bibitem[Chopin, 2002]{chopin:2002}
Chopin, N. (2002).
\newblock A sequential particle filter method for static models.
\newblock {\em Biometrika}, 89:539--552.

\bibitem[Cornuet and Beaumont, 2007]{cornuet:beaumont:2007}
Cornuet, J.-M. and Beaumont, M. (2007).
\newblock A note on the accuracy of {PAC}-likelihood with microsatellite data.
\newblock {\em Theoretical Population Biology}, 71:12--19.

\bibitem[Cornuet et~al., 2008]{cornuetetal:2008}
Cornuet, J.-M., Santos, F., Beaumont, M., Robert, C., Marin, J.-M., Balding,
  D., Guillemaud, T., and Estoup, A. (2008).
\newblock Infering population history with \emph{DIYABC}: a user-friendly
  approach to {A}pproximate {B}ayesian {C}omputation.
\newblock {\em Bioinformatics}, 24:2713--2719.

\bibitem[Del~Moral et~al., 2006]{delmoral:doucet:jasra:2006}
Del~Moral, P., Doucet, A., and Jasra, A. (2006).
\newblock Sequential {M}onte {C}arlo samplers.
\newblock {\em Journal of the Royal Statistical Society: Series B},
  68:411--436.

\bibitem[Douc et~al., 2007a]{douc:guillin:marin:robert:2007a}
Douc, R., Guillin, A., Marin, J.-M., and Robert, C. (2007a).
\newblock Convergence of adaptive mixtures of importance sampling schemes.
\newblock {\em Annals of Statistics}, 35:420--448.

\bibitem[Douc et~al., 2007b]{douc:guillin:marin:robert:2007b}
Douc, R., Guillin, A., Marin, J.-M., and Robert, C. (2007b).
\newblock Minimum variance importance sampling via {P}opulation {M}onte
  {C}arlo.
\newblock {\em ESAIM: Probability and Statistics}, 11:427--447.

\bibitem[Douc and Moulines, 2008]{douc:moulines:2008}
Douc, R. and Moulines, E. (2008).
\newblock Limit theorems for weighted samples with applications to sequential
  {M}onte {C}arlo methods.
\newblock {\em Annals of Statistics}, 36:2344--2376.

\bibitem[Doucet et~al., 2001]{doucet:defreitas:gordon:2001}
Doucet, A., {{de Freitas}}, N., and Gordon, N. (2001).
\newblock {\em Sequential {M}onte {C}arlo Methods in Practice}.
\newblock Springer-Verlag.

\bibitem[Doucet et~al., 2000]{doucet:godsill:andrieu:2001}
Doucet, A., Godsill, S., and Andrieu, C. (2000).
\newblock On sequential {M}onte {C}arlo sampling methods for {B}ayesian
  filtering.
\newblock {\em Statistics and Computing}, 10:197--208.

\bibitem[Gordon et~al., 1993]{gordon:salmon:smith:1993}
Gordon, N., Salmond, J., and Smith, A. (1993).
\newblock A novel approach to non-linear/non-{G}aussian {B}ayesian state
  estimation.
\newblock {\em IEEE Proceedings on Radar and Signal Processing}, 140:107--113.

\bibitem[Haario et~al., 1999]{haario:saksman:tamminen:1999}
Haario, H., Saksman, E., and Tamminen, J. (1999).
\newblock Adaptive proposal distribution for random walk {M}etropolis
  algorithm.
\newblock {\em Computational Statistics}, 14:375--395.

\bibitem[Haario et~al., 2001]{haario:saksman:tamminen:2001}
Haario, H., Saksman, E., and Tamminen, J. (2001).
\newblock An adaptive {M}etropolis algorithm.
\newblock {\em Bernoulli}, 7:223--242.

\bibitem[Hesterberg, 1995]{hesterberg:1995}
Hesterberg, T. (1995).
\newblock Weighted average importance sampling and defensive mixture
  distributions.
\newblock {\em Technometrics}, 37:185--194.

\bibitem[Hey and Nielsen, 2004]{hey:nielsen:2004}
Hey, J. and Nielsen, R. (2004).
\newblock Multilocus methods for estimating population sizes, migration rates
  and divergence time, with applications to the divergence of drosophila
  pseudoobscura and d.persimilis.
\newblock {\em Genetics}, 167:747--760.

\bibitem[Liu, 2001]{liu:2001}
Liu, J. (2001).
\newblock {\em {M}onte {C}arlo Strategies in Scientific Computing}.
\newblock Springer-Verlag.

\bibitem[Liu et~al., 2001]{liu:liang:wong:2001}
Liu, J., Liang, F., and Wong, W. (2001).
\newblock A theory of dynamic weighting in {M}onte {C}arlo computation.
\newblock {\em Journal of the American Statistical Association}, 96:561--573.

\bibitem[Nelder and Mead, 1965]{Nelder:Mead:1965}
Nelder, J. and Mead, R. (1965).
\newblock A simplex method for function minimization.
\newblock {\em Computer Journal}, 7:308--313.

\bibitem[Ortiz and Kaelbling, 2000]{ortiz:kaelbling:2000}
Ortiz, L. and Kaelbling, L. (2000).
\newblock Adaptive importance sampling for estimation in structured domains.
\newblock In {\em Proceedings of the Sixteenth Annual Conference on Uncertainty
  in Artificial Intelligence (UAI-2000)}, pages 446--454, San Francisco, CA.
  Morgan Kaufmann Publishers.

\bibitem[Owen and Zhou, 2000]{owen:zhou:2000}
Owen, A. and Zhou, Y. (2000).
\newblock Safe and effective importance sampling.
\newblock {\em Journal of the American Statistical Association}, 95:135--143.

\bibitem[Peel and McLachlan, 2000]{Peel:McLachlan:2000}
Peel, D. and McLachlan, G. (2000).
\newblock Robust mixture modelling using the $t$ distribution.
\newblock {\em Statistics and Computing}, 10:339--348.

\bibitem[Pennanen and Koivu, 2004]{pennanen:koivu:2004}
Pennanen, T. and Koivu, M. (2004).
\newblock An adaptive importance sampling technique.
\newblock In Niederreiter, H. and Talay, D., editors, {\em Monte Carlo and
  Quasi-Monte Carlo Methods 2004}, pages 443--455. Springer-Verlag.

\bibitem[Raftery and Bao, 2010]{raftery:bao:2010}
Raftery, A. and Bao, L. (2010).
\newblock Estimating and projecting trends in {HIV/AIDS} generalized epidemics
  using {I}ncremental {M}ixture {I}mportance {S}ampling.
\newblock {\em Biometrics}, 66:1162--1173.

\bibitem[Ripley, 1987]{ripley:1987}
Ripley, B.~D. (1987).
\newblock {\em Stochastic simulation}.
\newblock {John Wiley \& Sons Inc.}

\bibitem[Robert and Casella, 2004]{robert:casella:2004}
Robert, C. and Casella, G. (2004).
\newblock {\em {M}onte {C}arlo Statistical Methods}.
\newblock Springer-Verlag, second edition.

\bibitem[Rubinstein and Kroese, 2004]{rubinstein:kroese:2004}
Rubinstein, R. and Kroese, D. (2004).
\newblock {\em The cross-entropy method. A unified approach to combinatorial
  optimization, Monte-Carlo simulation, and machine learning}.
\newblock Springer-Verlag.

\bibitem[Sir\'en et~al., 2010]{siren:marttinen:corander:2010}
Sir\'en, J., Marttinen, P., and Corander, J. (2010).
\newblock Reconstructing population histories from single-nucleotide
  polymorphism data.
\newblock {\em Molecular Biology and Evolution}, 28:673--683.

\bibitem[Stephens and Donnelly, 2000]{stephens:donnelly:2000}
Stephens, M. and Donnelly, P. (2000).
\newblock Inference in molecular population genetics (with discussion).
\newblock {\em Journal of the Royal Statistical Society: Series B},
  62:605--655.

\bibitem[Veach and Guibas, 1995]{veach:guibas:1995}
Veach, E. and Guibas, L. (1995).
\newblock Optimally combining sampling techniques for {M}onte {C}arlo
  rendering.
\newblock In {\em SIGGRAPH '95 Conference Proceedings}, pages 419--428,
  Reading, MA. Addison-Wesley.

\end{thebibliography}

\end{document}